\documentclass[12pt]{article}
\usepackage[pdftex]{graphicx}
\usepackage{caption}
\usepackage{wrapfig}
\usepackage{latexsym}
\usepackage[pdftex]{color}
\usepackage{pifont}
\usepackage{epstopdf}
\usepackage{tikz}
\usepackage{mathtools}
\usetikzlibrary{decorations.pathmorphing,patterns}

\newcommand{\bmp}[2][t]{\begin{minipage}[#1]{#2}}
\newcommand{\emp}{\end{minipage}}

\usepackage{amsmath}
\usepackage{amssymb}
\usepackage{textcomp}
\usepackage{tcolorbox}
\usepackage[title]{appendix}

\usepackage{sidecap}

\usepackage[square,numbers,sort&compress]{natbib} 

\newcommand{\bblubox}[1]{\begin{tcolorbox}[colframe=blue!75!white,title=#1]}
\newcommand{\eblubox}{\end{tcolorbox}}

\begin{document}
\title{Multiple Modes of Motion for the Effectiveness\\
of Outer Hair Cells at High Frequencies}
\date{}
\author{Kuni H. Iwasa\\
\normalsize{NIDCD, National Institutes of Health}\\
\normalsize{Bethesda, MD 20892, USA}}

\maketitle

\abstract{
Outer hair cells (OHCs) are essential for the sensitivity and frequency specificity of the mammalian ear. To perform this function, OHCs need to amplify the motion of the basilar membrane (BM), which is much stiffer than themselves. If OHCs and the BM are components of a single oscillator, this impedance mismatch seriously limits the effectiveness of OHCs. However, the elaborated structure of the organ of Corti can support multiple modes of motion. Here, systems of two coupled oscillators are examined as the simplest models of the system with multiple modes of motion.  It is found that some of these model systems have conditions, under which an OHC can function as an effective amplifier, overcoming the impedance mismatch. The present examination suggests that the presence of multiple modes of motion is a key to the exquisite performance of the mammalian ear. }

\section*{Significance}
The mammalian ear depends on outer hair cells as the cochlear amplifier. To perform this biological function, soft outer hair cells need to amplifying the vibration of the much stiffer basilar membrane, overcoming the stiffness mismatch. The present paper shows that a system of two coupled oscillator, the simplest of more complex systems, allows effective transmission of the power generated by outer hair cells to the vibration of the basilar membrane. Therefore, the elaborate structure of the inner ear, which supports multiple modes of motion, must be a key to the exquisite performance of the mammalian ear. In addition, the properties of outer hair cells obtained from isolated cells preparations are compatible with their physiological function.

\section*{Introduction}
The mammalian hearing range may extend up to 100 kHz or beyond \cite{Vater2011}, depending on the species, quite remarkable for a biological system. Such function may call for a special rapid mechanism.  Indeed, it is shown that ``electromotility'' of outer hair cells (OHCs), a kind of piezoelectricity with which their cylindrical cell bodies are driven by the receptor potential, is essential for the sensitivity and frequency selectivity of the mammalian ear \cite{Dallos2008}. In addition, how OHCs are integrated into the rest of the cochlea is a critical issue, which requires extensive study to understand the mechanical basis of the mammalian hearing  \cite{Robles2001,deBoer2010}. 

Some of the earlier theories treated the activity of OHCs as an additional pressure applied to the basilar membrane (BM) \cite{deBoer1980} or impedance of the BM \cite{kimetal1980,z1991}. Some other treatments placed OHCs between two inertial masses, resulting in multiple degrees of freedom \cite{nk1986,ChadDimIws1996,Liu2015,Altoe2022,Rabbitt2023}. The goal of these theories has been to explain the tuning curve of the ear. 

Now, let us focus on the issue of the upper bound of auditory frequency. One of the difficulties of explaining the effectiveness of OHCs was the low-pass nature of their intrinsic electric circuit. This problem can be addressed by introducing piezoelectric resonance \cite{mh1994,Iwasa2017}. However, the model employed in the treatment assumed that soft OHCs and the stiff BM are components of the same oscillator. The upper limit obtained was $\sim$10 kHz, much higher than the roll-off frequency (up to several kHz) of the intrinsic electric circuit, but still not high enough for covering the auditory frequency \cite{Iwasa2017}.

A possible solution to the limitation imposed by impedance mismatch is that the OHCs and the BM are elements of separate oscillators, and those oscillators are coupled so that energy can be transferred between them. Multiple modes of motion appears to be consistent with recent experimental reports, including those with optical coherence tomography (OCT) \cite{Jawadi2016}. Those reports show complex motion of the inner ear, which depends on the stimulation intensity as well as frequency \cite{Cooper2018,Dong2018,Frost2022,He2022}.

Although recent studies have shown that the neuronal output of the cochlea is more closely correlated to the motion of the reticular lamina rather than that of the BM \cite{Ren2016a,Guinan2012}, BM motion is essential because it is the basis for the tonotopic map of the cochlea \cite{bekesy1960}. The objective of the present study is to examine the performance of simple cases of coupled oscillators, which retain the essential features of the cochlea, using the properties of OHCs determined from isolated cell preparations. 

We examine a set of two oscillators, heavy and light, coupled with either elastic or viscous element as the simplest examples of a system with multiple modes of motion.  At first, a single mode oscillator with an OHC is described to illustrate the effect of impedance mismatch. Then systems of two coupled oscillators are examined. The light oscillator (LO) incorporates an OHC and the heavy one (HO) includes the BM. Coupling is either elastic or viscous. The OHC is stimulated by either the motion of HO or that of LO.

The present treatment focuses on local energy balance, and intended to address the upper bound of the auditory frequency, even though the condition of local energy balance applies to any location, where traveling waves of a given frequency stops \cite{Wang2016}. 

\section*{Single mode of motion}

\begin{SCfigure}[1.6] 
\includegraphics[width=0.35\linewidth]{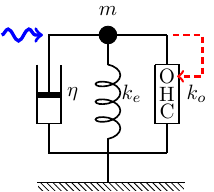} 
\caption{\small{A single oscillator model. The oscillator consists of mass $m$, an external elastic element with stiffness $k_e$, and a damper with drag coefficient $\eta$ and an outer hair cell (OHC), which responds to the movement of mass (dashed red arrow). This system is driven by a sinusoidal waveform with angular frequency $\omega$ (wavy blue arrow).}}
\label{fig:mech_schem1}
\end{SCfigure}

Let us start from two simple model oscillator systems, in which an OHC is incorporated (Fig.\ \ref{fig:mech_schem1}). The equation of motion can be formally written as
 \begin{eqnarray}
\left(m\frac{d^2}{dt^2}+\eta\frac{d}{dt}+k_o+k_e \right)X=F_\mathrm{OHC}+F,
\label{eq:singleX}
\end{eqnarray}
where $m$ is the mass, $\eta$ drag coefficient, $k_e$ the stiffness of the external elastic load, $k_o$ the material stiffness of the OHC, $X$ is the length of the OHC, and $k_e$ the stiffness of external load. $F$ is an external force and $F_\mathrm{OHC}$ is the force generated by the OHC.

OHC force  $F_\mathrm{OHC}$ is generated by the difference of mechanoelectric coupling free energy between the given moment and the values of for model illustrated by Fig.\ \ref{fig:mech_schem1} can be expressed by the free energy difference \cite{Iwasa2016,Iwasa2017,Iwasa2021}.

Consider the case, where a force $F$ applied has a periodic waveform with angular frequency $\omega$, we can write
\begin{subequations}
\begin{align}
V&=V_0+v\exp[i\omega t],\\
X&=X_0+x \exp[i\omega t], \\
F&=F_0+f \exp[i\omega t],
\end{align}
\end{subequations}
where $V$ is the receptor potential, which is generated by the mechanosensitivity of the hair bundle.

The derivation of the equation for the single oscillator model follows previous papers  \cite{Iwasa2016,Iwasa2017,Iwasa2021} as shown in Appendix. The basic assumption underlying the equation is that the driving force of the OHC at a given moment is due to the free energy difference from the equilibrium state for the given voltage and the force applied to the motile element of the OHC. 

The equation of motion can be expressed using changes in hair bundle resistance $\hat r$ instead of voltage changes $v$, as follows. See Appendix for details of the derivation.
\begin{align}
  (-\overline\omega^2+i\overline\omega/\overline\omega_\eta+1+\gamma B)x=i\frac{\gamma A}{\overline\omega}\; \hat r+\hat f,
  \label{eq:oneOHC}
\end{align}
where
\begin{subequations}
\label{eq:ABandf}
\begin{align}
A&=\frac{uk_o}{\omega_r(k_e+k_o)}\\
B&=u_a k/(k_e+k_o)+u_q, \\
\hat f&=f/(k_e+k_o),
\end{align}
\label{eq:AB}
\end{subequations}
and $\overline\omega(=\omega/\omega_r)$ is the frequency normalized to the mechanical resonance frequency $\omega_r=\sqrt{(k_e+k_o)/m}$, and $\overline\omega_\eta$ is normalized viscoelastic roll-off frequency. 

Quantities $u$, $u_a$, and $u_q$ represent hair bundle sensitivity and piezoelectricity of the OHC. More specifically,  $u$ represents mechanoelectric coupling, $u_a$ mechanical term, and $u_a$ electrical term, while all of them depend on hair bundle sensitivity. See Table \ref{tab:param_defs} for definitions (for derivations see Appendix). The quantity $\gamma$ represents relative voltage sensitivity of prestin, the motile membrane protein of OHCs \cite{zshlmd2000}. The maximal value is set to unity (i.e. $0\leq\gamma\leq 1$). It will be referred to as the ``operating point parameter.'' Notice also that both $A$ and $B$ depend on the ratio $k_e/k_o$ of external load to the intrinsic stiffness of the OHC.

The first term of the right-hand side of Eq.\ \ref{eq:oneOHC}, which contains $A$, works as an amplifying term, counteracting the drag term on the left-hand side. However, it is inversely proportional to the frequency $\omega$, unlike negative drag. The reason for this frequency dependence stems from the receptor potential, where capacitive current overwhelms resistive current at high frequencies. The parameter $B$ represents the contribution of the motile element to the stiffness of the cell body of the OHC.

\begin{table}[h]
\caption{Parameter definitions}
\begin{center}
\begin{tabular}{|c|p{4cm}||c|c|}
\hline \hline
symbol & definition & parameter & definition \\
\hline
$P_0$ & motor conformation$^\star$ & $\beta$ & $1/(k_BT)$ \\
$a$ & prestin displacement &  $\gamma$ & $ 4 P_0(1-P_0)$ \\
$\omega_r$ & resonance frequency & $u_a$ & $\beta a^2k_oN/4$  \\
$q$ & prestin charge& $u_b$ & $\beta a q N/4$  \\
$N$ & number of prestin & $u$ & $\beta i_0 a q N/(4C_0)$  \\
$C_0$ & cell capacitance & $u_q$ & $\beta q^2 N/(4C_0)$ \\
$i_0$ & resting current & $\kappa$ $^{\star\star}$ & $k_o/(k_o+k_e+k_c)$ \\
$k_o$ & cell stiffness & $A$ &  $(u/\omega_r)\; \kappa$ \\
$k_e$ & parallel elastic load & $B$ & $u_a(1-\kappa)+u_q$ \\
$k_c$ & coupling spring & $f_B$ & external force (amplitude)\\
$\omega_1$ & resonance frequency ratio & $\omega_\eta$ & $K/\eta$ \\
$\hat r$ & relative change of hair bundle resistance & & \\
\hline
\end{tabular}
\end{center}
\caption*{\small{Note: $k_B$ is Boltzmann's constant, $T$ the temperature. $^\star$ The motile function of OHCs is described by a two state model. $P_0$ is the fraction of one of the two states in a single cell at the operating point. $^{\star\star}$ The definition of $\kappa$ given here assumes the presence of $k_c$, the elastic element between the two oscillators. In the absence of such coupling, $k_c=0$. See Appendix for derivations. }}
\label{tab:param_defs}
\end{table}%

For a small amplitude $x$, $\hat r$ can be proportional to $x$. Thus, we may put $\hat r=g x$. This substitution leads to the equation of motion for the single mode model
\begin{align} \label{eq:single_mode}
  [-\overline\omega^2+i(\overline\omega/\overline\omega_\eta - \gamma gA/\overline\omega)+1+\gamma B]x=\hat f.
\end{align}
Notice the amplitude $x$ increases by the term with $A$, which counteracts the drag term.

The performance of the OHC in the system can be quantified by power gain $G(\omega)$, which is the ratio of power output to power input. Power output is in the form of viscous dissipation $\eta \,|dX/dt|^2$ because other the component is recovered during a cycle. Power input can be expressed as $\mathrm{Re}[F\cdot dX/dt]$. Thus, this quantity can be expressed by
\begin{align} \nonumber
G(\omega)&=\eta \omega^2|x|^2/|\omega f\; \mathrm{Im}[x] |\\
&=\eta\omega|x|/(f \sin \phi),
\label{eq:pGainX}
\end{align}
where $x$ is given by Eq.\ \ref{eq:single_mode}, $\mathrm{Re}[...]$ and $\mathrm{Im}[...]$ represent, respectively, real part and imaginary part, and $\phi$ is the absolute value of the phase angle of $x$ with respect to external force. In the absence of the OHC, i.e.\ $A=B=0$, we obtain $G(\omega)=1$ as expected.

\section*{Coupled oscillator models}
Consider a system with two harmonic oscillators, light and heavy. The light oscillator (LO) consists of an OHC and an elastic load $k_e$ and inertia $m$. The displacement of this oscillator is $X$. The heavy oscillator (HO) consists of the BM and inertia $M$, an elastic element with stiffness $K$, and viscous load with drag coefficient $\eta$. We can assume in general that the two oscillators are coupled by an elastic element with stiffness $k_c$ and a viscous element with drag coefficient $\eta_c$, and external force $F$ is applied to the HO (Figs.\ \ref{fig:two_H} and \ref{fig:two_L}). The motion of the two oscillators are described by,
\begin{subequations}\label{eq:DO_DO}
\begin{align}
\left(m \frac{d^2}{dt^2}+\eta_1 \frac{d}{dt}+k_e+k_o \right)X&=F_\mathrm{OHC}+\left(k_c+\eta_c\frac{d}{dt}\right)(Y-X)\\
\left(M \frac{d^2}{dt^2}+\eta_2 \frac{d}{dt}+K\right)Y&=\left(k_c+\eta_c\frac{d}{dt}\right)(X-Y)+F(t),
\end{align}
\label{eq:general_coupling}
\end{subequations}
\noindent where $F_\mathrm{OHC}$ is active force exerted by the OHC, being stimulated at its hair bundle.

An issue for this model is which oscillator includes the main drag with coefficient $\eta$. Here we assume that the main drag belongs to HO. If we assume that the main drag belong to LO, the OHC only amplifies the amplitude of LO and it cannot affect the oscillation of HO.

Classical analyses  show that energy transfer between two coupled oscillators is rather complex despite the apparent simplicity of the equations \cite{Morse_Ingard,Mercer1971}. Here we only examine only the response to continuous sinusoidal stimulation. 

\begin{subequations}
\begin{align}
F&=F_0+f _B\exp[i\omega t],\\
X&=X_0+x \exp[i\omega t],\\
Y&=Y_0+y \exp[i\omega t],
\end{align}
\end{subequations}
where $\omega$ is the angular frequency of stimulation. Notice here that notation $f_B$ is used for the amplitude of external force in the systems of coupled oscillators.
Now $F_\mathrm{OHC}$ can be expressed in a manner similar to in Eq.\ \ref{eq:single_mode} using $A$ and $B$. In the case of viscosity coupling these quantities are defined by Eq.\ \ref{eq:AB}. With elastic load coupling these quantifies are expressed by
\begin{subequations}
\label{eq:ABmod}
\begin{align}
A&=(u/\omega_r)\; k_0/(k_e+k_o+k_c) \\
B&=u_a(k_e+k_c)/(k_e+k_c+k_o)+u_q.
\end{align}
\end{subequations}

In general, bending of the OHC hair bundle that results in $\hat r$ may be proportional to $x$, the displacement of the LO as well as $y$, the displacement of the HO.  For simplicity, two extreme cases will be examined here:  $\hat r$ depends only on $y$ (HO-driven mode), or  only on $x$ (LO-driven mode). Coupling may have both viscous and elastic components. However, we assume coupling is purely elastic or purely viscous for simplicity.  

Power gain $G(\omega)$ is obtained analogous to Eq.\ \ref{eq:pGainX} for the simple oscillator,
\begin{align}
G(\omega)=\eta\omega |y|/(f_B \sin \varphi),
\label{eq:pGainY}
\end{align}
where $\varphi$ is the absolute value of the phase angle of $y$. Notice that amplifier gain here is determined by the amplitude $y$ of HO because both external force and drag work on HO.

\subsection*{HO-driven case} 

The assumptions that the main drag of the system belongs to HO, and that the dominant drag is due to the shearing motion in the subtectorial space, may lead to hair bundle sensitivity to HO motion (Fig.\ \ref{fig:two_H}), i.e. $\hat r=g Y$.

\subsubsection*{Viscous coupling}
HO has viscoelastic roll-off frequency $\omega_\eta$.  
With new parameters defined by $s=\!K/k$ and re-defined external force amplitude $f=f_B/K$,
the set of equations can be written as
\begin{subequations}
\label{eq:HOVeq}
\begin{align}\label{eq:y-driven-va}
[-(\overline\omega/\overline\omega_1)^2 +1+is\overline\omega/\overline\omega_c+\gamma B]x-i[\gamma Ag/\overline\omega+s\overline\omega/\overline\omega_c]y=0,\\ 
\label{eq:y-driven-vb}
-i\overline\omega/\overline\omega_c \;x+[-\overline\omega^2+i\overline\omega/\overline\omega_\eta+1+i\overline\omega/\overline\omega_c]y=f,
\end{align}
\end{subequations}
where $\omega_r$ is the resonance angular frequency of the HO, i.e.\ $\omega_r^2=K/M$ and the frequency is normalized to this resonance frequency. For example $\overline\omega=\omega/\omega_r$ and $\overline\omega_c=\omega_c/\omega_r$ with $\omega_c=K/\eta_c$. The quantity $\overline\omega_1$ is the ratio of the resonance frequency of the LO to that of HO. Notice that $f$ has the dimensionality of length. 

\subsubsection*{Elastic coupling}
Coupling introduced by an elastic element $k_c$ would elevate the resonance frequency of LO more than that of HO. Since energy transfer between the oscillators depends on the difference in the resonance frequencies of the two oscillators, an parameter $\omega_1$ is introduced to adjust the resonance frequency of LO.

\begin{SCfigure}[1.7]
\includegraphics[width=0.3\linewidth]{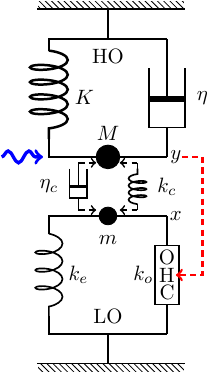} 
\caption{\small{Coupled oscillators, in which OHC is driven by the HO. The HO (top) consists of mass $M$, an elastic element with stiffness $K$, and a damper with drag coefficient $\eta$. The light oscillator (bottom) consists of mass $m$, an elastic element with stiffness $k_e$ an OHC, which responds to the movement of the HO (dashed red arrow). The HO is driven by a sinusoidal waveform with angular frequency $\omega$ (wavy blue arrow) and the two oscillators are coupled either by an elastic element with stiffness $k_c$ or by a damper with viscous coefficient $\eta_c$.}}
\label{fig:two_H}
\end{SCfigure}
\begin{subequations}
\label{eq:HOEeq}
\begin{align}\label{eq:y-driven-ea}
[-(\overline\omega/\overline\omega_1)^2 +1+cs+\gamma B]x-[i\gamma Ag/\overline\omega+cs]y=0,\\ \label{eq:y-driven-eb}
-cx+[-\overline\omega^2+i\overline\omega/\overline\omega_\eta+1+c]y=f,
\end{align}
\end{subequations}
\noindent where $c$ is defined by $c=\!k_c/K$. 

\subsection*{LO-driven case}
An alternative assumption regarding OHC stimulation is that the displacement $x$ of the LO affects the bending of OHC hair bundle, creating a direct feedback loop (Fig.\ \ref{fig:two_L}). 

\subsubsection*{Viscous coupling}
The set of equations of motion can be written as
\begin{subequations}
\label{eq:LOVeq}
\begin{align}\label{eq:x-driven-va}
[-(\overline\omega/\overline\omega_1)^2 -i\gamma Ag/\overline\omega+1+is\;\overline\omega/\overline\omega_c +\gamma B]x-i s\;\overline\omega/\overline\omega_c  y=0,\\ \label{eq:x-driven-vb}
-i\;\overline\omega/\overline\omega_c x+[-\overline\omega^2+i\overline\omega/\overline\omega_\eta+1+i\;\overline\omega/\overline\omega_c]y=f.
\end{align}
\end{subequations}
The phase relationship between $x$ and $y$ is determined by Eq.\ \ref{eq:x-driven-va}.

\subsubsection*{Elastic coupling}
The time dependent components follow the following equation
\begin{subequations}
\label{eq:LOEeq}
\begin{align}\label{eq:x-driven-a}
[-(\overline\omega/\overline\omega_1)^2 +1+cs+\gamma B-i\gamma Ag/\overline\omega]x-csy=0,\\ \label{eq:x-driven-b}
-cx+[-\overline\omega^2+i\overline\omega/\overline\omega_\eta+1+c]y=f,
\end{align}
\end{subequations}

\begin{SCfigure}[1.85]
\includegraphics[width=0.28\linewidth]{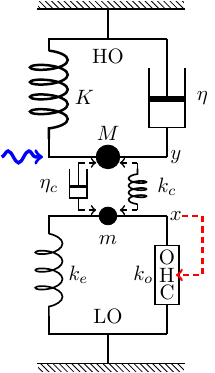} 
\caption{\small{LO-driven coupled oscillators. The HO consists of mass $M$, an elastic element with stiffness $K$, and a damper with drag coefficient $\eta$. The LO (bottom) consists of mass $m$, an elastic element with stiffness $k_e$ an OHC, which responds to the movement of the LO (dashed red arrow). The HO is driven by a sinusoidal waveform with angular frequency $\omega$ (wavy blue arrow) and the two oscillators are coupled either by an elastic element with stiffness $k_c$ or a dashpot with viscous coefficient $\eta_c$.}}
\label{fig:two_L}
\end{SCfigure}

\section*{Parameter values}

For numerical analysis of the performance of the coupled oscillators described, determining the values for parameters $A$ and $B$, OHC stiffness $k_o$ and BM stiffness $K$  is essential. For this purpose experimental values from guinea pigs near 20 kHz location are available. See Table \ref{tab:params20k}.

\subsection*{Cellular factors (20 kHz)}
Parameters $A$ and $B$ depend on the electromotility parameters $u$, $u_a$, and $u_q$ (See Table \ref{tab:param_defs}). For a 20 $\mu$m long cell, typical of the 10 to 20 kHz cochlear region, the linear capacitance is $C_0 = 8$ pF and $an = 1 \mu$m, which is 5\% of the resting cell length. Most in vitro experiments show the unitary motile charge of $q = 0.8 e$, where $e$ is the electronic charge. The membrane potential is near the optimal range for the motile element. The resting basolateral resistance is 7 M$\Omega$ and the resting membrane potential of $-50$ mV requires the resting apical resistance of 30 M$\Omega$. These values lead to $i_0 = 4$ nA. These parameter values are summarized in Table \ref{tab:params20k}

\begin{table}[h]
\caption{Parameter values at 20 kHz location}
\begin{center}
\begin{tabular}{c|c|c|c}
\hline
parameter & definition & value & source \\
\hline\hline
$C_0$ & structural capacitance of OHC & 8 pF &\\
$k_o$ & structural stiffness of OHC & 20 mN/m & \cite{ia1997} \\
$k_e$ & elastic load on OHC & adjustable & \\
$K$ & BM stiffness per OHC & 200/3 mN/m & see text \\
$s_o$ & ratio $K/k_o$ & 10 &  \\
$s$ & stiffness ratio $K/k$ & $\sim$5 & \\
$i_0$ & mechanoreceptor current & 4 nA & \cite{Johnson2011}\\
$\omega_\eta/\omega_r$ & viscous roll-off frequency&12.5 & see text \\
$g$ & hair bundle sensitivity & 1/(25 nm)& \cite{Nam2015,rrc1986} \\
\hline
\end{tabular}
\end{center}
\label{tab:params20k}
\end{table}%

The stiffness $k_o$ of a 20 $\mu$m long OHC, which may correspond to the same location, is about 20 mN/m, given the specific stiffness of 510 nN/m per unit strain for guinea pigs \cite{ia1997}. However, this value may have some uncertainty. The bottom part, up to 10 $\mu$m, of the OHC is held by the Deiters' cup. If this structure works as a damper \cite{nobili-mammano1993}, the length of elastic displacement is larger because the displacement includes the part within the cup. If, on the contrary, the structure is tight and rigid, not allowing slippage, the value of $k_o$ must be higher. 

\begin{SCfigure}[1.1]
\includegraphics[width=0.35\linewidth]{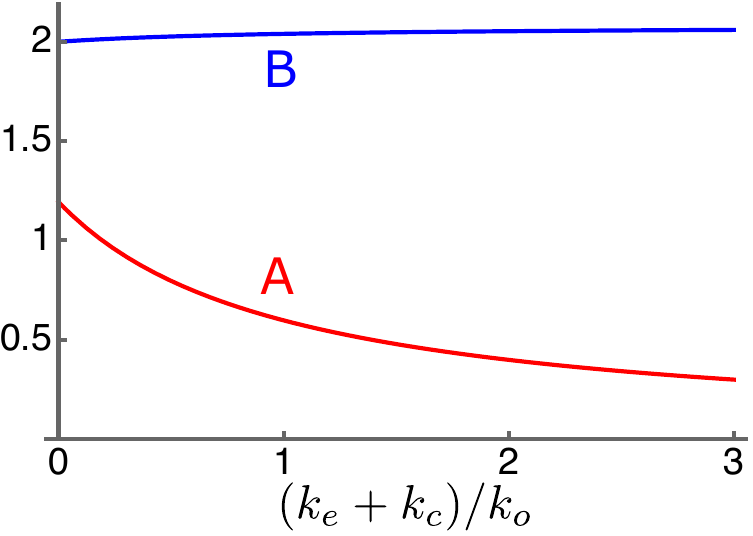}
\caption{\small{Dependence of parameters $A$ (red) and $B$ (blue) on the elastic load. The abscissa is $(k_e+k_c)/k_o$, the ratio of the external elastic load to the intrinsic stiffness of OHC.}}
\label{fig:AB}
\end{SCfigure}

The parameter values in Table \ref{tab:params20k} lead to a set of values for an OHC at 20kHz location:
\begin{align}
u/\omega_r(20k)=1.8, \quad u_a(20k)=0.08, \quad u_q(20k)=2.
\end{align}
Since $A=(u/\omega_r)\kappa$ and $B=u_a(1-\kappa)+u_q$ as defined earlier by Eq.\ \ref{eq:AB}, $B$ is approximately 2. It is not sensitive to the load ratio $\kappa$ because it $u_a$ is smaller than $u_q$ (Fig.\ \ref{fig:AB}).

\subsection*{Cochlear factors (20 kHz)}
Assume that the stiffness $K$ of HO is from the BM.
 At the location of 20 kHz best frequency in guinea pigs is 3 mm from the stapes according to the Greenwood function, assuming the length of the BM is 18.5 mm \cite{greenwood90,olson2012}. The stiffness of the location is about 0.21 N/m measured with a probe with 25 $\mu$m diameter tip \cite{Gummer1981}. This value is compatible with the ones obtained with a probe with 10 $\mu$m tip \cite{Miller1985}.   Thus, the stiffness ratio $s_o$ is $\sim$10.

The intrinsic mechanical resonance frequency of the location is somewhat uncertain because of the so-called ``half-octave shift'' \cite{Robles2001}. If the intrinsic mechanical resonance corresponds to the ``passive'' condition, the resonance frequency of the 20 kHz location is 14 kHz ($\omega_r=20/\sqrt{2}$) being a half octave lower. However, it could be the opposite because viscous damping brings the peak to a lower frequency if it is not counteracted.
 
A significant contribution to friction is expected from the gap between the tectorial membrane and the reticular lamina. The friction coefficient of this gap can be estimated by a formula $\mu S^2/d$, where $\mu$ is the viscosity of the fluid, $S$ the area of the gap per OHC, and $d$ the gap,  provided that the thickness of the boundary layer, which is $\sim 3.6\; \mu$m for 20 kHz \cite{Batchelor}, is greater than the gap. If we assume $S$ is $10\mu m\times 15\mu m$ and $d$ $1\mu m$, the friction coefficient is $1.2\times 10^{-7}$ N/m \cite{odi2003a,Iwasa2017}. With the resonance frequency of 14 kHz, the gap friction leads to a value $12.5$ for $\omega_\eta/\omega_r$. 

\subsection*{Extrapolation to higher frequencies}
Extrapolation to higher frequencies includes OHC parameters, the amplification parameter $A$ and the shift parameter $B$ as well as the stiffness ratio $s$, which depends on the stiffness of the BM.

Of the OHC parameters, $B$ does not depend heavily on the resonance frequency (Fig.\ \ref{fig:AB}) because it is dominated by the electrical term $u_q$ (See Eq.\ \ref{eq:ABandf}b). The amplification parameter $A$ decreases at higher frequency locations because it is inversely proportional to the resonance frequency $\omega_r$. However, such reduction is compensated by other factors. OHCs at higher frequency locations have larger resting current $i_0$ owing to larger hair bundle conductance and larger structural stiffness $k_o$ of the cell body owing to its shorter cell length (See Eq.\ \ref{eq:ABandf}a and Table \ref{tab:param_defs}).

At higher frequency locations, the stiffness ratio $s$ is expected to increase because the stiffness of the BM would increase more than the stiffness of OHCs. However, we proceed by assuming that the ratio $s$ remains the same. The reason is an uncertainty in the effective length of OHCs. The stiffness is inversely proportional to the effective length of the lateral membrane, which is harder to determine with shorter OHCs because the connectivity in Deiters' cup is ambiguous as described earlier. In addition, somewhat higher values for $s$ do not lead to qualitatively different results.

To examine high frequency performance aiming at 40kHz, twice higher than 20kHz, where parameter values are examined above, a numerical analysis is performed, assuming the following set of parameter vales:
\label{eq:assumedABs}
\begin{align}
u/\omega_r(40k)=1.8/2, \quad
B(40k)= 2, \quad
k/k_o=1, \quad
s=5, \quad \omega_\eta/\omega_r=10.
\label{used_set}
\end{align}
It is expected that the frequency of the location is somewhat higher than 40 kHz.

\section*{Single mode oscillator} 

\begin{SCfigure}[1.25]
\includegraphics[width=0.4\linewidth]{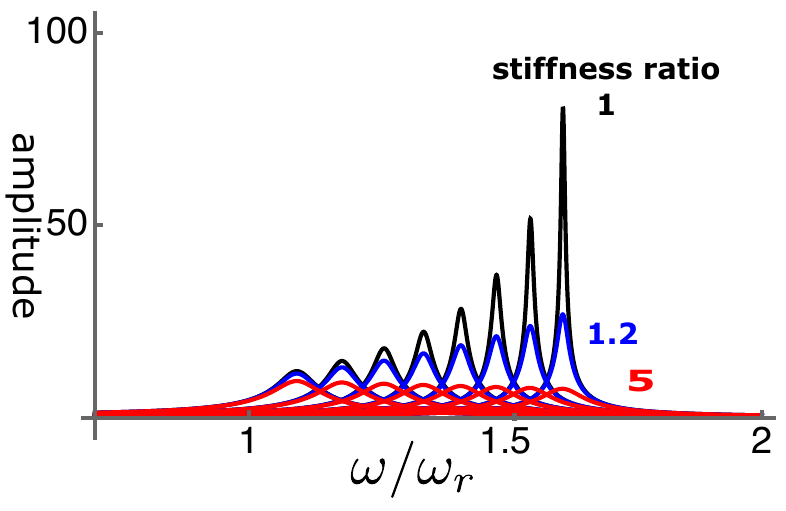}
\caption{\small{Amplitude of single mode oscillator plotted against frequency (normalized to $\omega_r$). Peak values are re-plotted in Fig.\ \ref{fig:SO}A. Stiffness ratio of load to OHC is 1, 1.2, and 5. $\gamma$ (operating point variable) runs from 0.03 (left) to 0.24 with increment of 0.03. The unit of the ordinate axis is $f(=F/K)$}.}
\label{fig:SOfreq}
\end{SCfigure}

\begin{SCfigure}[0.6]
\includegraphics[width=0.6\linewidth]{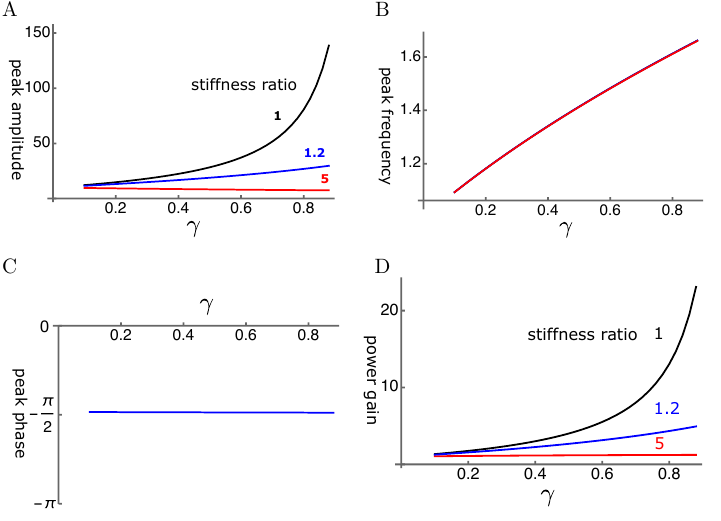} 
\caption{\small{A: Peak amplitude, 
B: Peak frequency (normalized to the resonance frequency), C: peak phase, and D: The ratio of power output to power input are respectively plotted against $\gamma$, the operating point variable.  The peak phase is $\pi/2$ in all cases.
}}
\label{fig:SO}
\end{SCfigure}

Let us start from the performance of the single mode oscillator before examining coupled oscillator models. The steady state amplitude is obtained by solving Eq.\ \ref{eq:single_mode}. The amplitude is normalized by the input $\hat f$ (Figs.\ \ref{fig:SOfreq}). As $\gamma$, the operating point variable of the OHC, increases, the peak amplitude shifts to higher frequencies independent of the elastic load (Fig. \ref{fig:SO}B). The peak height sharply increases with $\gamma$ if the elastic load is relatively small, i.e. the ratio $k_e/k_o$ is 1 or smaller (Fig.\ \ref{fig:SO}A).
However, the amplitude does not increase with $\gamma$ even if $k_e/k_o$ is as small as 1.2. With larger elastic load, such as $k_e/k_o\geq 5$, the peak amplitude does not increase at all.

The phase of the oscillator is delayed by $\pi/2$ as expected from the maximal amplitude of the damped oscillator (Fig.\ \ref{fig:SO}C). That makes the ratio of power output to power input similar to the amplitude of oscillation (Fig.\ \ref{fig:SO}D) because the power input insensitive to the magnitude of the drag. Thus, power gain is significantly attenuated by input impedance mismatch.

\section*{Coupled oscillators} 
In the following, we will examine if OHCs can be effective as amplifiers in coupled oscillators more than in a single mode oscillator. The coupled oscillator models have more parameters, which include stiffness ratio $s$ and resonance frequency ratio $\omega_1$ of the two oscillators, and coupling parameters (elastic element $c$ and viscous element $\eta_c$).

Introduction of elastic coupling introduces an additional elastic load to the OHC. This affects $A$ as well as $B$, which is not as sensitive as $A$. It also shifts the resonance frequency. Because this shift is much larger for LO, the resonance frequency of LO needs adjustment.

Conditions for large gain in amplitude are sought in all four cases: HO driven modes with elastic coupling and viscous coupling, and LO driven modes with elastic and viscous coupling. For each condition, a single example is given. These examples are typical ones but they do not necessarily show all shared features of the conditions.

\subsection*{HO-driven cases} 
It would be natural to start from HO-driven mode because we assumed that HO is associated with the dominant drag, which likely stems from the shear in the subtectorial space and this shear should stimulate the hair bundle of the OHC (Fig.\ \ref{fig:two_H}). 

\subsubsection*{Viscous coupling} 
The Deiters' cup that links an OHC with the BM via Deiters' cell could provide viscous coupling due to its morphology. For this reason, it is of interest to examine the effect of viscous coupling.

\begin{SCfigure}[1]
\includegraphics[width=0.5\linewidth]{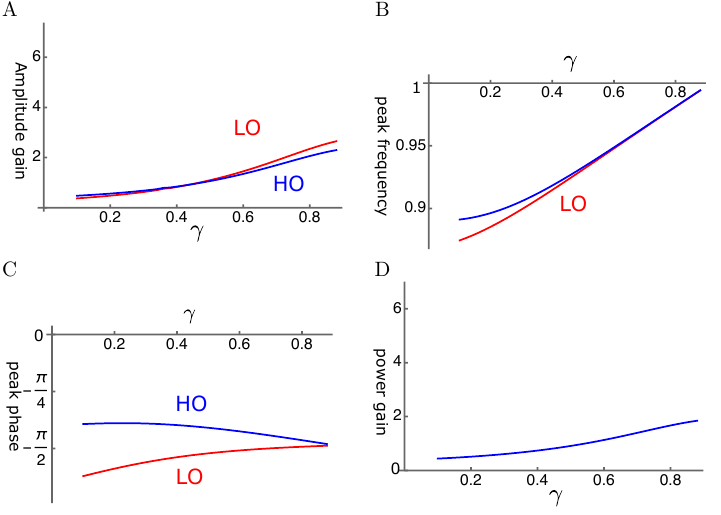} 
\caption{\small{HO-driven viscosity coupled oscillators (HOV). 
A: Amplitude gain over SO mode,
B: Peak frequency, and
C: Peak phase with respect to the external force
of each oscillator are plotted against $\gamma$, the operating point variable. LO (red) and HO (blue). 
D: Power input output ratio, which is closely related to HO.
The set of parameter values: $1/\overline\omega_c$=0.3, $\omega_1$=0.33 in Eq.\ \ref{eq:HOVeq}.
}}
\label{fig:HOV}
\end{SCfigure}

Even though viscous coupling of the oscillators can have amplifying effect, it does not appear to be so effective. Amplitude gain tends to be much smaller than that of elastic coupling (Fig.\ \ref{fig:HOV}A). Peak frequency of LO and that of HO are close to each other except for at small $\gamma$ (Fig.\ \ref{fig:HOV}B). The phase of HO is ahead of LO (Fig.\ \ref{fig:HOV}C). Since the force LO applies to HO is by $\pi/2$ ahead of the phase of LO, LO amplifies the motion of HO. The ratio of power output to input ratio is rather small (Fig.\ \ref{fig:HOV}D) as expected from the amplitude of HO.

\subsubsection*{Elastic coupling}
\begin{SCfigure}[1]
\includegraphics[width=0.5\linewidth]{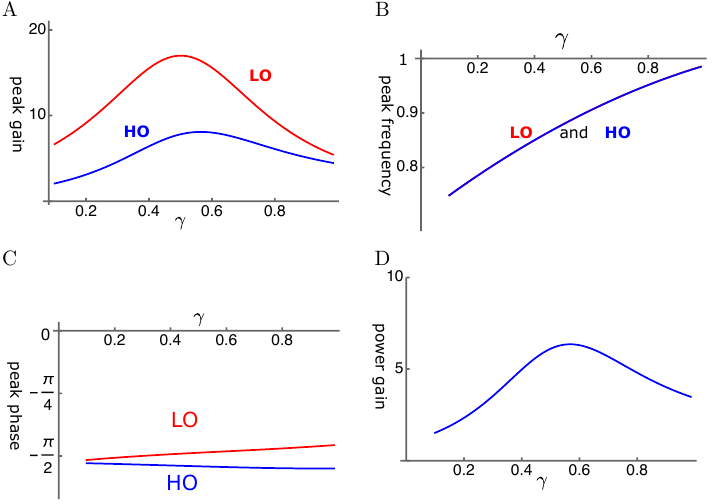} 
\caption{\small{HO-driven elastically coupled oscillators (HOE). 
A: Amplitude gain over SO mode,
B: Peak frequency, and
C: The phase with respect to the external force,
of each oscillator are plotted against $\gamma$. LO (red) and HO (blue). 
D: The ratio of power output to power input. 
The set of parameter values: $c$=0.2, $\omega_1$=0.29 in Eq.\ \ref{eq:HOEeq}.
}}
\label{fig:HOE}
\end{SCfigure}

Here, we assume that the coupling of the two oscillators is elastic. It is possible to find conditions, under which the amplitude of the LO reaches $\sim$20 fold larger than the amplitude of the single mode oscillator (Fig.\ \ref{fig:HOE}A). The abscissa is the operating point variable, which depends the voltage. It reaches unity at the optimal operating voltage. The peak  frequency is about the same for the two oscillators and increases with the operating point variable (Fig.\ \ref{fig:HOE}B). 

The phase advance LO with respect to HO is consistent with the amplifying role of LO. However, the phase difference is rather small (Fig.\ \ref{fig:HOE}C).
The ratio of power output to power input coincides with the amplitude maximum of HO and reached about 6 at the peak (Fig.\ \ref{fig:HOE}D).

\subsection*{LO-driven cases} 

Here we  assume that the OHC is primarily driven by the motion of LO rather than that of HO (Fig.\ \ref{fig:two_L}) to examine if the amplitude gain can be larger. The meaning of this mode will be discussed later. It turns out that LO-driven modes are more effective than HO-driven modes.

\subsubsection*{viscous coupling} 

\begin{SCfigure}
\includegraphics[width=0.5\linewidth]{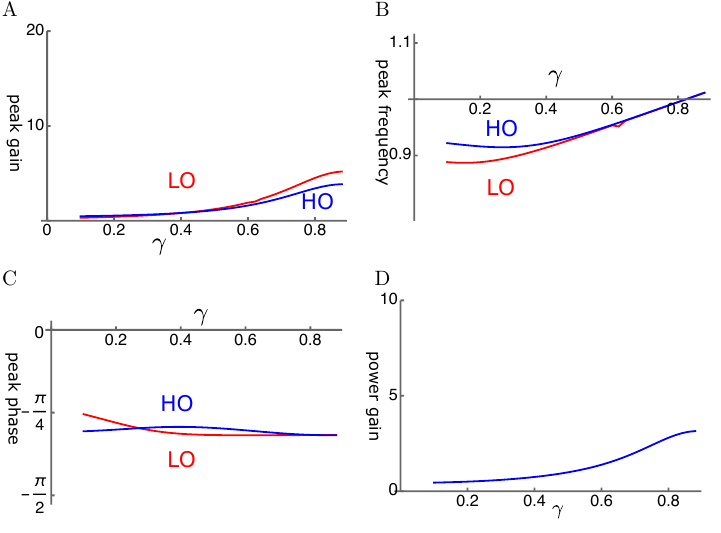} 
\caption{\small{LO-driven viscosity-coupled oscillator (LOE). 
A: Amplitude gain over SO mode,
B: Peak frequency, and
C: The phase with respect to the external force,
of each oscillator are plotted against $\gamma$. LO (red) and HO (blue). 
$\overline\eta_c=\eta_c\omega_r/K$.
D: Power input output ratio.
The set of parameter values: $1/\overline\omega_c$=0.2, $\omega_1$=0.36 in Eq.\ \ref{eq:LOVeq}.
}}
\label{fig:LOV}
\end{SCfigure}

The amplitude gain for both LO and HO increases with the operating point variable $\gamma$ (Fig.\ \ref{fig:LOV}A). HO has somewhat higher peak frequency than LO for small $\gamma$ (Fig.\ \ref{fig:LOV}B). The phases of HO and LO are both close to $\pi/4$ and HO tends to be ahead of LO except for where  $\gamma$ is small (Fig.\ \ref{fig:LOV}C). The power amplification is not very large, similar to the amplitude of HO. 

\subsubsection*{elastic coupling} 
\begin{figure}[h]
\begin{center}
\includegraphics[width=0.7\linewidth]{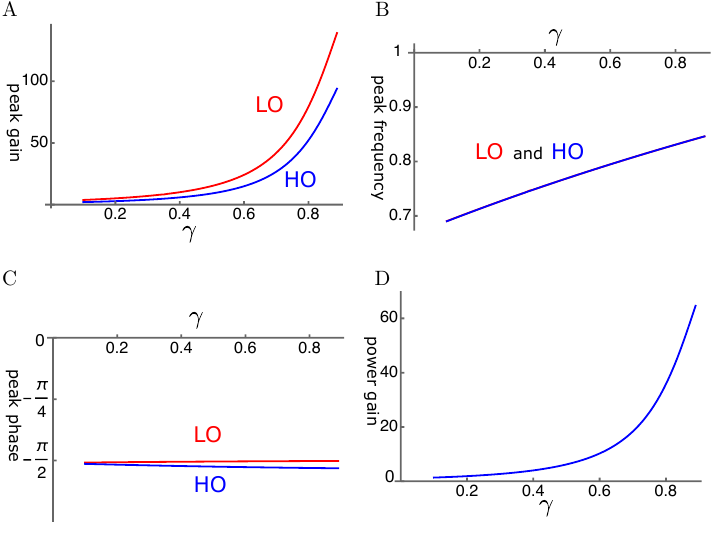} 
\end{center}
\caption{\small{An example of LO-driven elastically coupled oscillators (LOE).
A: Amplitude gain over SO mode,
B: Peak frequency, and
C: The phase with respect to the external force,
of each oscillator, is plotted against $\gamma$. LO (red) and HO (blue). 
D: The ratio of power output to power input. 
The set of parameter values: $c$=0.6, $\omega_1$=0.18 in Eq.\ \ref{eq:LOEeq}.
}}
\label{fig:LOE}
\end{figure}

An example is shown in Fig.\ \ref{fig:LOE}. 
The amplitude gain can exceed 100 fold for LO and 80 fold for HO (Fig.\ \ref{fig:LOE}A). These values are somewhat less in Fig.\ \ref{fig:LOE}A. The amplitude gain generally increases with the operating point variable $\gamma$ as expected. This increase can be monotonic as in Fig.\ \ref{fig:LOE}A. However, it can peak before $\gamma$ reaches unity, similar to HO-driven case, depending on the set of parameter values. 

The peak frequencies are similar for both oscillators (Fig.\ \ref{fig:LOE}B). The LO is ahead of HO in phase, indicating the amplifying role of the LO even though the difference is rather small (Fig.\ \ref{fig:LOE}C). The effect of the LO in affecting the amplitude of LO is quite large despite its much smaller mass and stiffness. 

The ratio of power output to input maximizes at a value of about 60 where the amplitude of HO is maximized (Fig.\ \ref{fig:LOE}D). 

\begin{SCfigure}[0.6]
\includegraphics[width=0.65\linewidth]{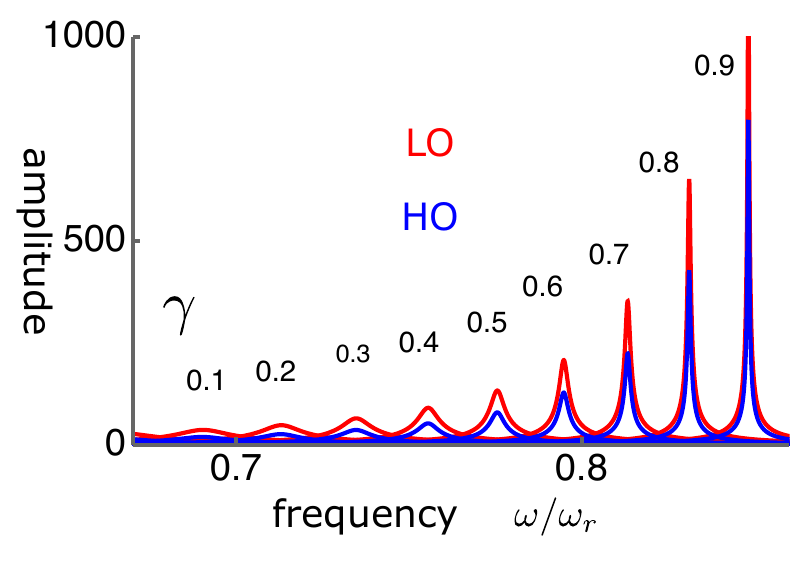}
\caption{\small{The amplitudes LO (red) and HO (blue) in LO-driven elastically coupled (LOE) oscillators. The peak values are plotted in Fig.\ \ref{fig:LOE}A. The abscissa is the frequency normalized to the resonance frequency $\omega/\omega_r$. The unit of the amplitude (ordinate axis) is $f/K$. The values of $\gamma$ are from 0.1(left) to 0.9(right) with an increment of 0.1.}}
\label{fig:LOEwf}
\end{SCfigure}

The frequency dependence shows quite sharp tuning together with the large amplitude gain (Fig.\ \ref{fig:LOEwf}).

\subsection*{Comparison of the cases} 
With the constraint of the given values of the parameters, it is possible to observe the major difference in the four major cases. In addition,  more generalized cases can be predicted by superposition of individual cases since our equations are linear.

The case of LO-stimulation with elastic coupling (LOE mode) is most effective in utilizing OHC.  It leads to about 60 fold power gain (Fig.\ \ref{fig:LOE}D) and amplitude gain of about 80 fold for HO (Fig.\ \ref{fig:LOE}D).  The case of HO-stimulation with elastic coupling (HOE mode) may lead to a relatively good amplitude gain for LO, even though it is less effective than LOE. In view of similar phase relationships of the oscillators in those two cases, a superposition of these two factors could be constructive.


\section*{Discussion}
The numerical examination performed here concerns only high frequency performance. At lower frequencies, such as 10 kHz, viscosity coupled models perform much better than at high frequencies (not shown). 

\subsection*{Input impedance mismatch}
The coupled oscillator models can make OHCs more effective by reducing the constraint of impedance mismatch. However, the performance of coupled oscillator models significantly varies depending on the type of coupling and how OHCs are stimulated.

More specifically, optimal amplification of this system by the OHC is achieved in the case where  coupling is elastic and the OHC is stimulated primarily by the oscillation of LO. Under such conditions, the gain is not impeded by a large stiffness mismatch between the OHC and the BM. 

\subsection*{Some characteristics}
Detailed properties of coupled oscillators vary depending on the set of parameter values. Under the conditions for larger amplifier gain, the following observations can be made.

\subsubsection*{Frequency dependence}
Strong coupling is required for making OHCs effective. In the case of elastic coupling, for example, the stiffness of the coupling element must be similar to that of the elastic element of LO. This condition appears to lead to another feature that the resonance frequency of LO is somewhat lower than that of HO. With increased OHC electromotility (increased $\gamma$), the resonance frequency tends to increase, reminiscent of the half-octave shift \cite{Ruggero1997,Ramamoorthy2012}. 

\subsubsection*{Stability and $\gamma$-dependence}
Amplitude gain of a coupled oscillator is in general an increasing function of $\gamma$, the the activity of the motile element in the OHC. Some of the cases (See Fig.\ \ref{fig:LOE}A) show monotonic increase with $\gamma$ similar to the single mode oscillator with low elastic load (See Fig.\ \ref{fig:SO}A). However, depending on the condition, it can peak before the operating point variable $\gamma$ reaches unity (See Figs.\ \ref{fig:HOE}A). 

Although such a behavior is not intuitive, it is compatible with a report that lowering prestin density in OHCs by 34\% does not have any reduction in the sensitivity of the ear \cite{Yamashita2012}. That is because both $A$ and $B$ are proportional to $N$, the number of motile units in an OHC (See Table \ref{tab:param_defs}). Thus, $N$ and $\gamma$ have the same effect on these parameters.

Amplitude gains are very sensitive to parameter values. They can be a smooth function of the operating point variable $\gamma$ as shown in the figures. They can show singular dependence on $\gamma$, indicating spontaneous oscillation. These properties could have bearing on otoacoustic emissions \cite{Kemp2010}. 

\subsection*{Structural implications}
The behavior of the equations examined must have a structural basis, including the nature of coupling elements, the effect of cell length, and movements in the subtectorial space.

\subsubsection*{Nature of coupling}
The cell body of OHCs is held by the reticular lamina at the apical end, forming tight junctions, and by Deiters' cup at the basal end. This structure introduces viscoelastic interaction with Deiters' cells \cite{Zhou2022}. Being held taut appears to indicate that OHCs are subjected to elastic load. However, Deiters' cap could provide viscous coupling \cite{nobili-mammano1993,Altoe2022}. The present analysis shows such a mode is not very effective to enhance the amplitude of LO at high frequencies. 

\subsubsection*{Cell length}
Short hair cells are more effective in the amplifying role having larger values of amplifying parameter $u$ because of their smaller $C_0$ and larger $k_o$. For this reason, basal cells, which operate high frequencies are short. For example, basal cells in both guinea pigs and mice are short ($\sim20\;\mu$m long), even though their length gradient is steeper for guinea pigs \cite{zajic1987}. 

\subsubsection*{Subtectorial drag and OHC stimulation}
The models examined in the present treatment assume that the main drag of the system is associated with HO. Without this assumption LO cannot influence the movement of HO. Such a condition could be realized if the main drag imposed on HO is outside the subtectorial gap.

However, the most likely source of the main drag of the system would be the shear in the subtectorial space between the tectorial membrane and the reticular lamina \cite{allen1980}. This picture indicates that the shear is associated with the motion of HO. If we accept that hair bundle bending is associated with this shear, the OHC should be stimulated by the displacement of HO. That leads to HO-stimulation models.

Can the hair bundle of the OHC be primarily stimulated by LO without incurring significant drag? Such a condition could be realized if LO moves the hair bundle in the direction perpendicular to the reticular lamina, resulting in effective bending of the hair bundle. However, the movement of LO should not incur significant drag. It would be possible that displaced water volume could be accommodated by local displacement of the TM owing to its pliability \cite{Teudt2014} and mechanical anisotropy \cite{Gavara2009,Masaki2009}. Then, the movement of LO does not result in viscous drag because it does not lead to fluid flow along the gap.

\subsection*{Speed of the motile element}
We assumed that prestin, the motile protein that drives OHCs, undergoes conformational transitions fast enough so that mechanical constraints determine the frequency dependence. This assumption is consistent with the experimental data on isometric force generation by OHC \cite{fhg1999} and current noise spectrum of OHC membrane \cite{Dong2000}. It is also in line with a recent analysis that movement of organ of Corti measured with OCT is consistent with the cycle-by-cycle force application \cite{Dewey2021}.  

However, the frequency of conformational changes must have an upper bound.
Recent repots that the roll-off frequencies of the voltage-dependent component of OHC membrane capacitance suggest 30 kHz, \cite{Santos2023} higher than older values of up to 20 kHz \cite{Santos-Sacchi2018,joe2019}. Those gating frequencies reported could reflect extracellular factors, such as viscoelastic process, of their experimental configuration. 

The present study provides a new perspective: with a large gain obtained by coupled oscillators, this issue is not so important. With a finite gating frequency $\omega_g$, the amplitude is attenuated by a factor $1/(1+\omega_g/\omega)$ \cite{Iwasa2004}. If the gating frequency is 20 kHz, this attenuation factor is 1/3 at 40 kHz, a small fraction of the expected gain. 

\subsection*{Implication to macroscopic models}
A good number of macroscopic models of cochlear mechanics \cite{deBoer2010} assume that OHCs apply force directly on the BM. Such models ignore issues resulting from microscopic structure, such as low-pass characteristics of intrinsic cellular electric circuit and impedance mismatch.

Interestingly, the present analysis of coupled oscillator models shows a resemblance to the assumptions of those macroscopic models: The amplifier gain is not seriously constrained by the impedance mismatch. Both the amplitude and the phase of the two oscillators are not so large.

\section*{Conclusions}

Large impedance mismatch between the BM and an OHC impedes energy transmission from the softer OHC to the oscillation of the stiffer BM if the BM and the OHC are in the same oscillator. However, if these elements are incorporated in separate oscillators that are coupled, a significant improvement in the efficiency of energy transmission can be achieved.

Among the modes of motion examined, the system of elastically coupled with LO-stimulated OHC is most effective in utilizing the OHC as the amplifier. Under optimal conditions, both amplitudes and phases of these oscillators are close because these oscillators are strongly coupled. 

The models examined here are the simplest possible cases. The real ear would be much more complex and can have more modes of motion. Nonetheless, these simple model systems could provide some insight into the working of the real system. It is likely that multiple modes of motion supported by the complexity of the organ of Corti is essential for the performance of the mammalian ear by making OHCs effective as the amplifier. 

\section*{Author Contributions}
The single author performed all, including conceptualization, development of the equations, computation using Mathematica, and composing the paper.

\section*{Acknowledgments}
I thank Drs.\  Richard Chadwick and Catherine Weisz  of NIDCD for helpful comments.  Dr.\ Chadwick pointed out the significance of Eq.\ 28 of Wang et al \cite{Wang2016}. I also appreciated help by Drs.\ Matthew Kelley and Inna Belyantseva for reading the manuscript. This research was supported in part by the Intramural Research Program of the NIH, NIDCD. 

\section*{Declaration of interests}
The author declares no competing interests.

\bibliography{/Users/kuni/Dropbox/wip/bib/ohc,../bib/vibr} 
\numberwithin{equation}{section} 
\numberwithin{figure}{section}
\renewcommand{\theequation}{A\arabic{equation}}
\renewcommand{\thefigure}{A\arabic{figure}}

\appendices 

\section{Single oscillator with OHC} \label{apx:OHC} 
Here we consider a case, in which an OHC incorporated into a system with mechanical resonance as illustrated in Fig.\ \ref{fig:mech_schem}. Described first is the case of electrical stimulation. Then the treatment is extended to mechanical stimulation. 

\subsection*{Electrical stimulation}
The description largely follows the one-dimensional model for electrical stimulation \cite{Iwasa2017}, rather than the more realistic cylindrical model \cite{Iwasa2021} for simplicity because the descriptions of OHCs by these two models are consistent.

\subsubsection*{Motile element} 
Assume that the motile element in the lateral membrane of the OHC have two states, the long state $L$ and the short state $S$, and $P$ is the fraction of the motile element in state $L$. The natural length of the cell is $X_0+aNP$, where $a$ is the contribution of a single motile element to the cell length associated with conformational change from $S$ to $L$. This conformational change accompanies movement of a charge $q$ across the membrane. The number of the motile elements in the cell is $N$. In the equilibrium condition, $P$ is determined by the free energy $G$, which is the sum of an electrical term $qV$ and a mechanical term $aF_e$, where $V$ is the membrane potential (or voltage) and $F_e$ the force applied to the cell due to mechanoelectric coupling \cite{i1994,i1990}.

\begin{SCfigure}[0.9] 
\includegraphics[width=0.5\linewidth]{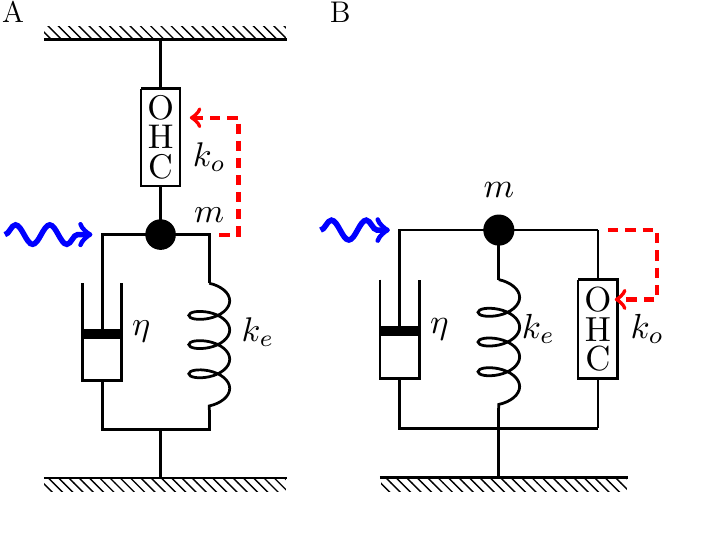} 
\caption{Two simple oscillators, A and B, each of which consists of an OHC with intrinsic stiffness $k_o$,  elastic load $k_e$ and viscous load $\eta$. The OHC responds to the movement of mass (dashed red arrow). The equations developed for A \cite{Iwasa2016,Iwasa2021} are equally applicable to B for electrical stimulation.  }
\label{fig:mech_schem}
\end{SCfigure}

Now put mechanical load, including an external elastic element $k_e$, a viscous element $\eta$, and a mass $m$, on an OHC with intrinsic stiffness $k_o$  (Fig.\ \ref{fig:mech_schem}). Assume that this system is initially in equilibrium, and then change the membrane potential by $\Delta V$. The motile element undergoes conformational change $\Delta P$. The electrical component of the free energy change is $q\Delta V$. The resulting mechanical displacement $\Delta X_p=aN\Delta P$ generates force $\Delta F_e$ due to the external elastic load. This force, in turn, produces an elastic displacement $\Delta X_e$ in the OHC, resulting in the net displacement $\Delta X=aN\Delta P-\Delta X_e$ on the OHC as well as the external elastic element $K$. Thus, $\Delta F_e=k_e \Delta X=k_o \Delta X_e$. These relationships leads to $ \Delta X=aN\Delta P\;k_o/(k_o+k_e)$ and
\begin{align}
 \Delta F_e=aN\Delta P\; k_ek_o/ (k_o+k_e).
 \label{eq:DeltaF}
\end{align}
Thus, the total free energy is $ q\Delta V+a\Delta F_e$ with $\Delta F_e$ given by Eq.\ \ref{eq:DeltaF}. 

We proceed by assuming that the free energy of the motile element at any given moment is determined by the given values of voltage $V$ and the mechanical strain $X$ of that moment, exactly the same as in the static case as described above. Then the variable $P$ of the motile element changes toward its equilibrium value $P_\infty$, which is given by the Boltzmann function
\begin{align}
P_\infty=\exp[-\beta\Delta G]/(1+\exp[-\beta\Delta G]),
\end{align}
with
\begin{align}
 \Delta G=G_0+q\Delta V+aN \Delta P\; k_ek_o/ (k_o+k_e),
\end{align}
and $\beta=1/(k_BT)$ with Boltzmann's constant $k_B$ and the temperature $T$,
and the constant term is represented by $G_0$. The difference $P-P_\infty$ drives the system.

\subsubsection*{Equation of motion} 
The equation of motion of the system can be formally expressed by
\begin{align}
 m \; d^2X/dt^2+\eta dX/dt=k_o(X_\infty-X),
 \label{eq:x_eqn}
\end{align}
where $X_\infty=aNP_\infty\; k_o/(k_o+K)$ is the displacement that corresponds to equilibrium. Here $m$ is the mass, $\eta$ drag coefficient, and $F$ external force. The inertia term can be justified if the system is not far from equilibrium \cite{Iwasa2021}.
This equation can be expresses using variable $P$
\begin{align}
m \; d^2P/dt^2+\eta dP/dt=(k_o+k_e)(P_\infty-P).
 \label{eq:p_eqn}
\end{align}
The amplitude $x$ of displacement is related to $p$ with
\begin{align}
  x=aNp\;k_o/(k_o+k_e).
  \label{eq:x_p}
\end{align}

Now consider the response to a sinusoidal voltage waveform of small amplitude $v$ and angular frequency $\omega$. Let $p$ the corresponding small amplitude of $P$.
\begin{align*}
V=V_0+v\exp[i\omega t], \quad
P=P_0+p\exp[i\omega t].
\end{align*}
Then the equation of motion (\ref{eq:p_eqn}) is transformed into
\begin{align}
 p=\frac{-\beta\gamma q/4}{-\overline\omega^2+i\overline\omega/\overline\omega_\eta+1+u_a k_e/(k_o+k_e)}\;v,
\label{eq:p}
\end{align}
where $\overline\omega(=\omega/\omega_r)$ is the frequency normalized to the mechanical resonance frequency $\omega_r=\sqrt{(k_o+k_e)/m}$, $\overline\omega_\eta$ is normalized viscoelastic roll-off frequency, $\gamma=4 P_0(1-P_0)$, and 
\begin{align}
u_a=\beta\gamma a^2k_oN/4. 
\label{eq:ua}
\end{align}
The coefficients $\gamma$ and $\alpha$ originate from the expansion of the exponential term of $P_\infty$.

\subsubsection*{Response to hair bundle stimulation}
The effect of hair bundle resistance $R_a$ on the membrane potential $V$ can be expressed
\begin{align}
 (e_{ec}-V)/R_a=(V-e_K)/R_m+C_0\; dV/dt-Nq\; dP/dt,
 \label{eq:Ra}
\end{align}
where $e_{ec}$ is the endocochlear potential, $e_K$ is the resting potential of OHC, which is primarily determined by K$^+$ conductance, and $R_a$ hair bundle conductance. The last term in the right-hand-side of the equation is due to the  change of the motile mechanism.

If we assume that stimulation at the hair bundle is periodic with angular frequency $\omega$, introducing the time independent component $R_0$ and the relative amplitude $\hat r$ of the hair bundle resistance $R_a$, we obtain
\begin{align}
 -i_0 \hat r=(\sigma+i\omega C_0)v-i\omega Nqp.
 \label{eq:rec_Pot}
\end{align}
Here $i_0=(e_{ec}-e_K)/(R_0+R_m)$ is  the steady state current and $\sigma=1/R_0+1/R_m$ the steady state conductance, which can be dropped for high frequency stimulation because it is smaller than $\omega C_0$.

From the transformed equation of motion Eq.\ \ref{eq:p} and equation for receptor potential Eq.\ \ref{eq:rec_Pot}, $p$ can be expressed as a linear function of $\hat r$.
\begin{align}
 p(-\overline\omega^2+i\overline\omega/\overline\omega_\eta+1+B)=i\frac{\beta\gamma i_0q}{\omega C_0}\; \hat r,
 \label{eq:p_eqn}
\end{align}
with
\begin{align}
B&=u_a k/(k_o+k_e)+u_q, \quad \mathrm{and} \\
u_q&=\beta\gamma Nq^2/(4C_0), 
\label{eq:uq}
\end{align}
which is the ratio of prestin capacitance at the zero-frequency limit to the structural capacitance. (appears in both 2017 paper and 2021 paper.) The equation shows that the damping term $1/\overline\omega_\eta=\omega_r/\omega_\eta$ must be small not to be overdamped. Notice also that $|p|$ is inversely proportional to $C_0$.

Eq.\ \ref{eq:p_eqn}, which is described by the variable $p$, can be rewritten by using $x$ as 
\begin{align}
  x(-\overline\omega^2+i\overline\omega/\overline\omega_\eta+1+\gamma B)=i\frac{\gamma A}{\overline\omega} \hat r,
  \label{eq:singleOHC}
\end{align}
where $A$ is defined by
\begin{align}
A&=\frac{i_ou}{\omega_r}\;\frac{k_o}{k_o+k_e} \quad\mathrm{with}\\
u&=\beta\gamma aqN/(4C_0).
\label{eq:u}
\end{align}

Notice that Eqs.\ \ref{eq:u}, \ref{eq:ua}, and \ref{eq:uq} are, respectively, definitions of $u$, $u_a$, and $u_q$, which appear in the main text.

\subsection*{Mechanical stimulation}
The equation of motion in the presence of an external force $F$ can be formally given by
\begin{align}
 m \; d^2X/dt^2+\eta dX/dt=k_o(X_\infty-X)+F.
 \label{eq:x_eqn}
\end{align}
If the motile mechanism in OHCs is as sensitive to applied force as to voltage changes, force stimulation requires a treatment that is symmetric to voltage stimulation. However, OHCs are a semi-piezoelectric resonator rather than a full piezoelectric resonator because he indirect effect of displacement via hair bundle conductance is more than an oder of magnitude greater than the direct effect of applied displacement on OHCs \cite{odi2003a}.

For this reason, the equation of motion can be expressed simply adding the external force as an additional term to the equation of motion developed for voltage stimulation. This applies to both models in Fig.\ \ref{fig:mech_schem}. We may expect Model B responds to an external force similar to Model A under the condition $k_e\lesssim k_o$, but such a condition is not needed because of the high sensitivity of the hair bundle.

Thus, we arrive the equation in the following form in both cases:
\begin{align} \label{eq:single_mode_app}
  [-\overline\omega^2+i(\overline\omega/\overline\omega_\eta - g\gamma A/\overline\omega)+1+\gamma B]x=\hat f,
\end{align}
if the change in the hair bundle resistance $\hat r$ can be expressed $gx$ and $\hat f=f/(k_e+k_o)$.
This equation is Eq.\ \ref{eq:single_mode} in the main text.

\end{document}